\documentclass[12pt,final]{article}
\usepackage{graphicx}
\usepackage{amssymb}
\usepackage{epsfig}
\title{Quasi-exactly solvable periodic potentials with three known eigenstates}
\author{O.~Voznyak\\
  {\small Ivan Franko Lviv National University,} \\
  {\small Department of Theoretical Physics,} \\
  {\small 12 Drahomanov Str., 79005 Lviv, Ukraine} \\
  {\small E-mail: ulex@ktf.franko.lviv.ua} }

\begin{document}
\hyphenation{su-per-po-ten-ti-al su-per-po-ten-ti-als}
 \maketitle

\begin{abstract}
Supersymmetric method of the constructing well-like quasi exactly
solvable (QES) potentials with three known eigenstates has been
extended to the case of periodic potentials. The explicit examples
are presented. New QES potential with two known eigenstates has
been obtained.
\end{abstract}

\section{Introduction}

Description of the electron's motion on a lattice has been
investigated for a long time as a central problem of the condensed
matter physics. Such quantum problem is reduced to the solving of
the Schr\"{o}dinger equation with some model potential which is
periodic often. Therefore, the periodic quantum mechanics problems
remain at the investigation's focus up to now.

The general properties of the solutions of Schr\"{o}dinger
equation with periodic potential energy are described by the
oscillation theorem \cite{HillsTheorem}. Energy spectrum of the
periodic potential has band structure, i.e. eigenvalues belong to
the allowed bands (energy bands) $[E_0,E_1],[E_1^{'},E_2],...$.
The wave functions are the Bloch functions, which are bounded and
extended on the full real axe
\begin{equation}
\psi(x+L)=\exp^{ikL} \psi(x),
\end{equation}
where $L$ is potential period and $k$ is a so-called
quasi-momentum. The limits of the energy bands are given by the
equation $kL=\{0,\pi\}$, and the wave functions, which belong to
the limiting energy values, satisfy the condition
$\psi(x+L)=\pm\psi(x)$. These energy values and wave functions are
often called the eigenvalues and the eigenfunctions of the
described above problem.

Oscillation theorem claims that in the case of periodic potentials
the eigenfunctions, which belong to the limits of the energy bands
and are arranged in the energy of the increasing order $E_0\leq
E_1\leq E_{1'} \leq E_2 \leq E_{2'} \leq E_3 ...$, are the
periodic functions with the period $L, 2L, 2L, L, L, 2L, 2L, ...$
and have $0,1,1,2,2,3,3,...$ nodes in the interval $L$
respectively.

Despite long term investigations, there is rather a limited number
of exactly solvable periodic potentials even in one dimension. The
classical examples are the Kronig-Penney model potential
\cite{KronigPenney} or Lam\'{e}'s potentials \cite{Lame}.

Because of limited number of the exactly solvable potentials,
recently much attention has been given to the quasi exactly
solvable (QES) potentials for which a finite number of the energy
levels and the corresponding wave functions are known explicitly.
A general treatment of the quasi exact solvability has been
introduced by Turbiner and Ushveridze \cite{Ushveridze}. The class
of QES trigonometric potentials was presented in \cite{Turbiner}.
In \cite{Turbiner2} it was shown that the Lam\'{e} equation is a
peculiar example of QES systems. The authors of the paper
\cite{QES1} considered a family of spectral equation which extends
those of \cite{Turbiner2}. Authors of \cite{QHJ} has applied
quantum Hamilton-Jacobi formalism to the QES periodic potentials.
In the latest paper \cite{Unified} an unified treatment of quasi
exactly solvable potentials was proposed.

The powerful tool for studying the problem of exact solvability of
the Schr\"{o}dinger equation is the supersymmetric (SUSY) quantum
mechanic introduced by Witten \cite{Witten} (for a review of SUSY
quantum mechanics see \cite{Cooper}). The SUSY method for
constructing QES potentials was used for the first time in
\cite{SUSY1} - \cite{SUSY3}. The idea of this method starts from
some initial QES potential with $n+1$ known eigenstates and using
the properties of the unbroken supersymmetry to obtain the SUSY
partner potential, which is a new QES one with $n$ know
eigenstates.

In \cite{QES4}-\cite{QES6} using the formalism of SUSY quantum
mechanics a large number of new solvable and QES periodic
potential was proposed. It is worth mentioning recent paper
\cite{QES7}, where the highest order SUSY transformations was
applied for studying periodic potential.

In recent Tkachuk's papers \cite{Tk1}-\cite{TkKuliy} a new SUSY
method for constructing of the QES potentials with two and three
known eigenstates has been proposed. This method does not require
knowledge of the initial QES potential in order to generate a new
QES one. Within the frame of this method QES potentials has been
obtained for which the explicit form of the energy levels and the
wave functions of the ground and the excited states can be found.
After the paper \cite{Dolya} by Dolya and Zaslavskii, where they
showed how to generate QES potentials with arbitrary two known
eigenstates without resorting to the SUSY quantum mechanics, SUSY
method has been extended \cite{Tk3} for constructing QES
potentials with arbitrary two known eigenstates. In our recent
works using the SUSY method periodic \cite{Tk4} and disordered
\cite{Tk5} QES potentials were obtained.

In the present paper using the results of previous study
\cite{Tk1}-\cite{Tk5} we extend Tkachuk's SUSY method for
constructing QES periodic potentials with three known eigenstates.

\section{The Witten model of SUSY quantum mechanics}

Witten model of supersymmetric quantum mechanics is a quantum
mechanics of the matrix Hamiltonian
\begin{equation}
H = \left(\begin{array}{lll}
{H_+} & {0} \cr
{0}   & {H_-}
\end{array}
\right), \label{SUSY_hamiltonian}
\end{equation}
where Hamiltonians
\begin{equation}
H_{\pm}=-{1\over 2}{d^2\over dx^2}+V_\pm(x)=B^{\mp}B^{\pm} \label{factorization}
\end{equation}
are supersymmetric partners and
\begin{equation}
B^{\pm} = {1 \over \sqrt{2}} \Big(\mp {d \over dx} + W(x)\Big).
\label{Bpm}
\end{equation}
Here $\hbar = m = 1$ units are used. Function $W(x)$ is referred to as
superpotential, $V_\pm(x)$ are so-called supersymmetric partner potentials
\begin{equation}
2V_{\pm}(x) = W^2(x)\pm W'(x). \label{VWbound}
\end{equation}

Energy spectrum of the supersymmetric partners $H_+$ and $H_-$ is
identical except for zero-energy ground state which exists in the
case of the unbroken supersymmetry. This leads to twofold
degeneracy of the energy spectrum of $H$, except for the unique
zero-energy ground state. Only one of the Hamiltonians $H_\pm$ has
zero-energy eigenvalue. We shall use the convention that the
zero-energy eigenstate belongs to $H_-$
\begin{equation}
\left\{\begin{array}{lll}
{E^-_{n+1}} & {=} & {E^+_n} \cr
{E^-_0} & {=} & {0} \label{spectr}
\end{array}\right.,
\end{equation}
where $n=0,1,2,...$. The wave functions of the supersymmetric partners $H_\pm$
are related by the supersymmetric transformations
\begin{equation}
\left\{\begin{array}{lll}
{\psi^-_{n+1}(x)} & {=} & {{1\over\sqrt{E^+_n}}B^+\psi^+_n(x)}    \cr
{\psi^+_n(x)} & {=} & {{1\over\sqrt{E^-_{n+1}}}B^-\psi^-_{n+1}(x)} \label{SUSYtransform}
\end{array}\right..
\end{equation}

Due to the factorization $H_-=B^+B^-$, we can find solution of the
Schr\"{o}dinger problem for the eigenstate with zero energy
\begin{equation}
H_- \psi_0^-(x) = E_0 \psi_0^-(x) = 0.
\end{equation}
It is easy to see that
\begin{equation}
\psi^-_0(x)=C_0^-\exp{(- \int W(x) dx)}, \label{psi0}
\end{equation}
where $C_0^-$ is an arbitrary constant.

In the present paper we shall consider the systems on the full
real axe $-\infty < x < \infty$ with periodic superpotential. The
periodic superpotential $W(x+L)=W(x)$ leads to the periodic
potential energy $V_\pm(x+L)=V_\pm$, which results in the bounded
and extended eigenfunction. A satisfactory condition for the
existence of periodic eigenfunctions, written in the terms of the
SUSY quantum mechanics, is
\begin{equation}
\int_0^LW(x)=0.
\label{extended_wave_function_cond}
\end{equation}
In \cite{SUSYQM-p1,SUSYQM-p2} a detailed analysis of the SUSY
quantum mechanics was made for this case.

\section{SUSY constructing QES potentials}

We shall study the Hamiltonian $H_-$ with the potential energy
\begin{equation}
V_-(x) = W_0^2(x) - W_0'(x), \label{V_minus}
\end{equation}
the ground state of which is given by (\ref{psi0}).

Let us consider Hamiltonian $H_+$ which is the SUSY partner of
Hamiltonian $H_-$ . If we calculate the ground state of $H_+$ we
immediately find the first excited state of $H_-$ using the
degeneracy of the spectrum of SUSY Hamiltonian and SUSY
transformations (\ref{SUSYtransform}). In order to calculate the
ground state of $H_+$ let us rewrite Hamiltonian in the following
form
\begin{equation}
H_+=H_-^{(1)}+\epsilon, \qquad \epsilon > 0 \label{Hplus_Hminus}
\end{equation}
where
\begin{eqnarray}
H_-^{(1)}=B_1^+B_1^-, \\
B_1^\pm={1\over \sqrt{2}} \Big(\mp {d\over dx}+W_1(x) \Big),
\end{eqnarray}
and $W_1(x)$ is a some new function. Note that $\epsilon$ is the
energy of the ground state of $H_+$ since $H_-^{(1)}$ has
zero-energy ground state.

The ground state wave function of $H_+$ with the energy $E =
\epsilon$ is also zero energy wavefunction of $H_-^{(1)}$ and it
satisfies the equation
\begin{equation}
B_1^-\psi_0^+(x)=0.
\end{equation}
The solution of this equation is
\begin{equation}
\psi_0^+(x) = C_0^+ \exp\Big(-\int W_1(x)dx\Big),
\end{equation}
where $C_0^+$ is an arbitrary constant.

Using the SUSY transformation (\ref{SUSYtransform}) we can
calculate the wavefunction of the first excited state of $H_-$.
Repeating the described procedure for $H_-^{(1)}$ we can obtain
the second excited state for $H_-$ and so on. This procedure is
well known in the SUSY quantum mechanics (see review \cite{Cooper}
for example). The wavefunctions and corresponding energy levels
read
\begin{equation}
\left\{\begin{array}{lll}
{\psi_n^-(x)} & {=} & {C_n^- B_0^+...B_{n-2}^+B_{n-1}^+\exp\Big(-\int W_n(x)dx \Big)} \cr
{E_n^-} & {=} & {\sum_{i=0}^{n-1}\epsilon_i}
\end{array}\right.,
\label{wave_function_general} \end{equation} where $n=1,2,...,N$;
$\epsilon_0=\epsilon$, $B_0^\pm=B^\pm$, $W_0(x)=W(x)$, $C_n^-$ are
an arbitrary constants. Operators $B_n^\pm$ are given by
(\ref{Bpm}) with the superpotentials $W_n(x)$.

Equation (\ref{Hplus_Hminus}) rewritten for N steps
\begin{equation}
H_+^{(n)}=H_-^{(n+1)} + \epsilon_n,
\end{equation}
where $n=0,1,...,N-1$, leads to the set of equations for superpotentials
\begin{equation}
W_n^2(x)+W_n'(x)=W_{n+1}^2(x)-W'_{n+1}(x)+2\epsilon_n,
\label{set_of_W}
\end{equation}
where $n=0,1,...,N-1$.

Unfortunately, each of the equations in (\ref{set_of_W}) are the
Rikatti equation, which can not be solved in the general case.
Previously this set of equations was solved in special cases of
shape-invariant potentials \cite{Gendenstein} and self-similar
potentials for arbitrary $N$ (see review \cite{deSouza}). For
$N=1$ in the context of parasupersymmetric quantum mechanics one
can obtain a general solution of (\ref{set_of_W}) without
restricting ourselves to shape-invariant and self-similar
potentials \cite{Beckers}. In recent papers \cite{Tk1}-\cite{Tk5}
a solution of (\ref{set_of_W}) for $N=1$ and $N=2$ in order to
obtain non-singular QES potentials with two and three known
eigenstates respectively has been constructed.

Let us write set of equations (\ref{set_of_W}) for the case $N=2$ in the explicit form
\begin{equation}\left\{\begin{array}{l}
{W_0^2(x)+W_0'(x)=W_1^2(x)-W_1'(x)+2\epsilon_0} \cr
{W_1^2(x)+W_1'(x)=W_2^2(x)-W_2'(x)+2\epsilon_1} \cr
\end{array}\right..\label{set_of_3W}\end{equation}

It is convenient to introduce new functions
\begin{equation}\begin{array}{ll}
{
\left\{\begin{array}{l}
{W_+(x) = W_1(x) + W_0(x)} \cr
{W_-(x) = W_1(x) - W_0(x)} \cr
\end{array}\right.
}

{
\left\{\begin{array}{l}
{\tilde{W}_+(x) = W_2(x) + W_1(x)} \cr
{\tilde{W}_-(x) = W_2(x) - W_1(x)} \cr
\end{array}\right.,
}
\end{array}\label{Wplus_Wminus}\end{equation}
then superpotentials can be rewritten in the following form
\begin{equation}\begin{array}{ll}
{
\left\{\begin{array}{l}
{2W_0(x) = W_+(x) - W_-(x)} \cr
{2W_1(x) = W_+(x) + W_-(x)} \cr
\end{array}\right.
}

{
\left\{\begin{array}{l}
{2W_1(x) = \tilde{W}_+(x) - \tilde{W}_-(x)} \cr
{2W_2(x) = \tilde{W}_+(x) + \tilde{W}_-(x)} \cr
\end{array}\right..
}
\end{array}\label{W_through_Wplus_Wminus}\end{equation}
In the terms of new functions (\ref{Wplus_Wminus}) the set of
equations (\ref{set_of_3W}) read as follows
\begin{equation}\left\{\begin{array}{lll}
{W_+'(x)} {=} {W_-(x)W_+(x)+2\epsilon_0} \cr
{\tilde{W}_+'(x)} {=} {\tilde{W}_-(x)\tilde{W}_+(x)+2\epsilon_1} \cr
\end{array}\right. . \label{new_set_of_3W}\end{equation}

Note, that there are two terms for the $W_1(x)$ in the equations
(\ref{W_through_Wplus_Wminus}) with respect to $W_-(x)$ and with
respect to $\tilde{W}_+(x)$. This gives us a possibility to obtain
relation between $W_+(x)$ and $\tilde{W}_+(x)$
\begin{equation}
W_+(x) + {W_+'(x)-2\epsilon_0 \over W_+(x)} =
\tilde{W}_+(x) - {\tilde{W}_+'(x)-2\epsilon_1 \over \tilde{W}_+(x)}, \label{Step1}
\end{equation}
here (\ref{new_set_of_3W}) are used. It is easy to rewrite this
equation as follows
\begin{eqnarray}
W_+(x)\tilde{W}_+(x)[\tilde{W}_+(x)-W_+(x)]-[W_+(x)\tilde{W}_+(x)]' + \cr \nonumber
+ 2[\epsilon_1W_+(x)+\epsilon_0\tilde{W}_+(x)]=0,
\end{eqnarray}
or
\begin{equation}
U(x)\Big({U(x)\over W_+(x)} - W_+(x)\Big) - U'(x) +
2\Big(\epsilon_1W_+(x)+\epsilon_0 {U(x) \over W_+(x)}\Big) = 0,
\end{equation}
where we have introduced a new function
\begin{equation}
U(x) = W_+(x) \tilde{W}_+(x). \label{U_definition}
\end{equation}
We arrive again to the Riccati equation with respect to $U(x)$. On
the other hand, this is an algebraic equation with respect to
$W_+(x)$, which can be solved explicitly
\begin{equation}\left\{
\begin{array}{lll}
{W_+(x)} & {=} & {{2U(x)(U(x)+2\epsilon_0) \over U'(x)(1 + R(x))}}  \cr
{\tilde{W}_+(x)} & {=} & {{U'(x)(1+R(x)) \over 2(U(x) + 2\epsilon_0)}} \label{W_plus_tilde}
\end{array}\right.,
\end{equation}
where
\begin{eqnarray}
\mathfrak{R}(x) = 1 + 4
{U(x)(U(x)+2\epsilon_0)(U(x)-2\epsilon_1)\over U'(x)^2}, \quad
R(x) = \pm \sqrt{\mathfrak{R}(x)}. \label{square_root}
\end{eqnarray}
The square root $\mathfrak{R}(x)$ is a positively defined value, while the function $R(x)$
can be chosen in the form of $\mathfrak{R}(x)$ or $-\mathfrak{R}(x)$
within different intervals separated by zeros of the function $\mathfrak{R}(x)$.

Thus, we can start from an arbitrary function $U(x)$ to construct
the functions $W_+(x)$ and $\tilde{W}_+(x)$ given by
(\ref{W_plus_tilde}). Using (\ref{W_through_Wplus_Wminus}) we
obtain three consequent superpotentials
\begin{equation}\left\{\begin{array}{lll}
W_0(x) &=& {1\over 2} \left(W_+(x)-{W'_+(x)-2\epsilon_0\over
W_+(x)}\right) \\
W_1(x) &=& {1\over2} \left(W_+(x)+{W'_+(x)-2\epsilon_0\over
W_+(x)}\right) \\
W_2(x) &=& {1\over2} \left(\tilde{W}_+(x) + {\tilde{W}'_+(x) -
2\epsilon_1\over \tilde{W}_+(x)}\right)
\end{array}\right..\end{equation}
Then because of (\ref{wave_function_general}), we can find the
wavefunctions of three explicitly known eigenstates of the
Hamiltonian $H_-$
\begin{equation}\left\{\begin{array}{lll}
\psi^-_0(x) &=& C_0^-e^{- \int W(x) dx}\\
\psi_1^-(x) &=& C_1^-W_+(x)e^{- \int W_1(x) dx} \\
\psi_2^-(x) &=&
C_2^-\Big((W_0(x)+W_2(x))\tilde{W}_+(x)-\tilde{W}_+'(x) \Big) e^{-
\int W_2(x) dx} \label{solution-}
\end{array}\right.\end{equation}
where energy values are $E^-_0 = 0$, $E_1^-=\epsilon_0$,
$E_2^-=\epsilon_0 + \epsilon_1$ and potential energy
\begin{equation}
V_-(x) = {1\over 2} (W_0(x)^2-W_0'(x)). \label{V-}
\end{equation}

Simultaneously we can find the wave function of two explicitly
known eigenstates of the Hamiltonian $H_+$
\begin{equation}\left\{\begin{array}{lll}
\psi^+_1(x) &=& B_0^- \psi_1^-(x) \\
\psi^+_2(x) &=& B_0^- \psi_2^-(x) \label{solution+}
\end{array}\right.\end{equation}
with energy values $E_1^+ = \epsilon_0$, $E_2^+ = \epsilon_0 +
\epsilon_1$ and potential energy
\begin{equation}
V_-(x) = {1\over 2} (W_0(x)^2+W_0'(x)). \label{V+}
\end{equation}

Note that obtained terms for the superpotentials, potentials and
wave functions allow existence of two different solutions
depending on the selected sign before the square root
$\pm\sqrt{R(x)}$ in the $W_+(x)$ and $\tilde{W}_+(x)$ definitions.
Here and later we shall distinguish solutions which were obtained
for different signs, by superscript in the parenthesis after the
function designation, for example $Y(x)^{(+)}$. We will denote as
$Y(x)$ solutions which are identical for different signs
$Y(x)^{(+)}=Y(x)^{(-)}$.

Choosing different generating functions $U(x)$ we will obtain
different QES potentials (\ref{V-}) with three explicitly known
eigenstates (\ref{solution-}) and QES potentials (\ref{V+}) with
two explicitly known eigenstates (\ref{solution+}). Of course,
function $U(x)$ must satisfy some conditions to provide physical
solutions of the Schr\"{o}dinger equation.

The main obvious condition imposed on the function $U(x)$ is a
positivity of the expression under the square root
(\ref{square_root})
\begin{equation}
1+ {4U(x)(U(x)+2\epsilon_0)(U(x)-2\epsilon_1) \over U'(x)^2} \geq
0
\end{equation}
on the all periodicity interval.

Another set of restrictions imposed on function $U(x)$ appears due
to the requirement of the non-singularity of resulting potential
$V_-(x)$. The full analysis of the properties of superpotential
$W_0(x)$ which provides non-singular potential $V_-(x)$ and wave
functions $\psi_0^-(x)$, $\psi_1^-(x)$ in the terms of function
$W_+(x)$ was done in \cite{Tk3}-\cite{Tk5} for the case of the
quasi exactly solvable potentials with two exactly known
eigenstates. Below we extend this analysis to the case of the
quasi exactly solvable potential with three exactly known
eigenstates.

As we can see from the superpotentials $W_0(x)$, $W_1(x)$ and
$W_2(x)$ definitions, potential $V_-(x)$ can have poles at the
points $x_0$ where $W_+(x_0)=0$ or $\tilde{W}_+(x_0)=0$.
Fortunately, such poles can be removed when \cite{Tk3}
\begin{equation}\left\{\begin{array}{lll}
W_+'(x_0) &=& \pm 2\epsilon_0, \\
\tilde{W}_+'(x_0) &=& \pm 2\epsilon_1. \label{W_plus_d}
\end{array}\right.\end{equation}

Besides, potential energy $V_-(x)$ can have poles at the points of
singularity $x_\infty$ of the function $W_+(x)$. As it was shown
in \cite{Tk5}, if function $W_+(x)$ at the singularity points
$x_\infty$ has the behavior
\begin{equation}
W_+(x)=const+{-1\over x-x_\infty}+o(x-x_\infty),
\label{W_plus_pole}\end{equation} or
\begin{equation}
W_+(x)={-3\over x-x_\infty}+o(x-x_\infty) \label{W_plus_pole_2},
\end{equation}
obtained potential energy and wave functions will be continuous
functions at the points $x_\infty$.

To provide bounded and extended wave functions $\psi_0^-(x)$,
$\psi_1^-(x)$, $\psi_2^-(x)$ the conditions
(\ref{extended_wave_function_cond}) should be satisfied
\begin{equation}\left\{\begin{array}{lll}
  \int_0^L W_0(x) dx &=& 0 \\
  \int_0^L W_1(x) dx &=& 0 \\
  \int_0^L W_2(x) dx &=& 0 \\
\end{array}\right..\label{periodicity_condition}\end{equation}
These conditions are satisfied in the simplest way if the
corresponding superpotenials $W_0(x)$, $W_1(x)$, $W_2(x)$ are odd
function with regard to the middle of the periodicity interval
$x_m$. To obtain odd superpotentials it is enough to expect the
odd behavior of the function $W_+(x)$.

Let us choose the $U(x)$ as even function with regard to the
middle of the periodicity interval $x_m$. Then, if we apply
solutions with the different signs before square root to the parts
of the periodicity interval from the left and from the right of
$x_m$, $W_+(x)$ will be odd function. It is easy to see if we
rewrite term (\ref{W_plus_tilde}) for the $W_+(x)$ as follows
\begin{equation}
W_+(x)={2U(x)(U+2\epsilon_0)\over
U'(x)\pm\sqrt{U'(x)^2+4U(x)(U(x)+2\epsilon_0)(U(x)-2\epsilon_1)}}.
\end{equation}

Application of the solutions with the different signs leads to the
finite breaks of the function $W_+(x)$. These breaks can be
removed if the value of the function $W_+(x)$ will tend to zero
both from the left and right direction.

Thus, to provide existence of the bounded extended wave functions,
$U(x)$ should be even function with regard to the middle of the
periodicity interval $x_m$, and obtained function $W_+(x)$ should
have zero at the point $x_m$. Function $W_+(x)$ can have zeros at
the points, where $U(x)=0$, and, since $U(x)$ should be even
function with regard to $x_m$, function $U(x)$ can have at the
point $x_m$ zero of the even-order only.

Of course, function $U(x)$ can have zeros at the other points of
the periodicity interval too. Let us analyze in details the
behavior of the superpotentials, potential energy and the wave
functions in the vicinity of the $U(x)$ zeros.

Let the function $U(x)$ have the first-order zeros at the points
$x_0^a$
\begin{equation}
  U(x)=U'(x_o^a)(x-x_o^a)+{1\over
  2}U''(x_o^a)(x-x_o^a)^2+o(x-x_o^a)^3.
\end{equation}

Then behavior of the functions $W_+(x)$, $\tilde{W}_+(x)$ in the
vicinity of the points $x_0^a$ will be as follows
\begin{equation}\left\{\begin{array}{lll}
  W_+(x)^{(+)} &=& 2\epsilon_0(x-x_0^a)+o(x-x_0^a)^2 \\
  W_+(x)^{(-)} &=& {U'(x_0^a)\over 2\epsilon_1}+o(x-x_0^a) \\
  \tilde{W}_+(x)^{(+)} &=& {U'(x_0^a)\over 2\epsilon_0}+o(x-x_0^a) \\
  \tilde{W}_+(x)^{(-)} &=& 2\epsilon_1(x-x_0^a)+o(x-x_0^a)^2
\end{array}\right..\end{equation}
It is easy to see that functions $W_+(x)$ and $\tilde{W}_+(x)$ at
the points $x_0^a$ will have non-zero values or will have zeros
which satisfy (\ref{W_plus_d}).

Superpotentials $W_0(x)$, $W_1(x)$, $W_2(x)$ will be the following
\begin{equation}\left\{\begin{array}{lll}
  W_0(x)^{(\pm)} &=& A_0^{(\pm)} + o(x-x_0^a) \\
  W_1(x)^{(\pm)} &=& A_1^{(\pm)} + o(x-x_0^a) \\
  W_2(x)^{(\pm)} &=& A_2^{(\pm)} + o(x-x_0^a)
\end{array}\right.,\end{equation}
where
\begin{equation}\left\{\begin{array}{lll}
  A_0^{(+)} &=& -A_1^{(+)} = -{8\epsilon_0^2 \epsilon_1 +U'(x_0^a)^2-\epsilon_0U''(x_0^a)
  \over 2\epsilon_0 U'(x_0^a)} \\
  A_1^{(-)} &=& -A_2^{(-)} = {U'(x_0^a)^2+\epsilon_1(U''(x_0^a)-8\epsilon_0\epsilon_1)
  \over 2\epsilon_1 U'(x_0^a)} \\
  A_0^{(-)} &=& -A_2^{(+)} = -{U''(x_0^a)-8\epsilon_0\epsilon_1\over 2U'(x_0^a)} \\
\end{array}\right..\end{equation}
Obtained potential will be regular function too
\begin{equation}
  V_-(x)^{(\pm)} = \alpha_-^{(\pm)} + o(x-x_0^a),
\end{equation}
where
\begin{equation}\left\{\begin{array}{lll}
\alpha_-^{(-)} &=& -{64\epsilon_0^2\epsilon_1^2 + 8 \epsilon_1
U'(x_0^a)^2 + U''(x_0^a)^2
-16\epsilon_0(U'(x_0^a)^2+\epsilon_1U''(x_0^a))
-2U'(x_0^a)U^{(3)}(x_0^a)\over
8U'(x_0^a)^2} \\
\alpha_-^{(+)} &=&
-3\alpha_-^{(-)}+4\epsilon_0+{U^{(3)}(x_0^a)\over 2U'(x_0^a)}
\end{array}\right..\end{equation}
The wave functions $\psi_0^-(x)$, $\psi_1^-(x)$, $\psi_2^-(x)$
will read as follows
\begin{equation}\left\{\begin{array}{lll}
  \psi_0^-(x) = 1 + o(x-x_0^a) \\
  \psi_1^-(x)^{(+)} = 2\epsilon_0(x-x_0^a) + o(x-x_0^a)^2 \\
  \psi_1^-(x)^{(-)} = {U'(x_0^a)\over 2\epsilon_1}+o(x-x_0^a)  \\
  \psi_2^-(x) = -2\epsilon_1 + o(x-x_0^a)
\end{array}\right..\end{equation}
Thus, at the points $x_0^a$, where function $U(x)$ has first-order
zeros, potential $V_-(x)$ and wave functions $\psi_0^-(x)$,
$\psi_1^-(x)$, $\psi_2^-(x)$ will be continuous functions, and
wave function $\psi_1^-(x)$ can have simple zeros at the points
$x_0^a$ depending on the selected sign before the square root.

Let us consider potential $V_+(x)$, which is the supersymmetric
partner of the obtained potential $V_-(x)$. Potential $V_+(x)$
will be regular function in the vicinity of the points $x_0^a$ too
\begin{equation}\left\{\begin{array}{lll}
V_+(x)^{(\pm)} = \alpha_+^{(\pm)} + o(x-x_0^a) \\
\alpha_+^{(+)}=\alpha_-^{(-)}+2\epsilon_1+{U'(x_0^a)^2-2\epsilon_0U''(x_0^a)\over
4\epsilon_0^2} \\
\alpha_+^{(-)}=\alpha_-^{(+)}-2\epsilon_1
\end{array}\right.\end{equation}
with continuous wave functions
\begin{equation}\left\{\begin{array}{lll}
  \psi_1^+(x) = \sqrt{2} \epsilon_0 + o(x-x_0^a) \\
  \psi_2^+(x)^{(+)} = 2\sqrt{2}\epsilon_1(\epsilon_0+\epsilon_1)(x-x_0^a) + o(x-x_0^a)^2
  \\
  \psi_2^+(x)^{(-)} = {(\epsilon_0 + \epsilon_1) U'(x_0^a) \over \sqrt{2}\epsilon_0} +
  o(x-x_0^a)
\end{array}\right..\end{equation}
Depending on the selected sign before the square root wave
function $\psi_2^+(x)$ can have nodes at the points $x_0^a$.

Now let the function $U(x)$ have second-order zeros at the points
$x_0^b$
\begin{equation}
  U(x)={1\over 2}U''(x_0^b)(x-x_o^b)^2+{1\over
  6}U^{(3)}(x_o^b)(x-x_o^b)^3+o(x-x_o^b)^4,
\end{equation}
then behavior of the functions $W_+(x)$ and $\tilde{W}_+(x)$ will
be as follows
\begin{equation}\left\{\begin{array}{lll}
\label{psi_nodes_analogy}
W_+(x) = 2\epsilon_0(x-x_0^b)+o(x-x_0^b)^{3/2} \\
\tilde{W}_+(x) = 2\epsilon_1(x-x_0^b)+o(x-x_0^b)^{3/2}
\end{array}\right..\end{equation}
i.e. at the points $x_0^b$ functions $W_+(x)$ and $\tilde{W}_+(x)$
will have zeros. Keeping in mind (\ref{W_plus_d}), it is easy to
obtain the following coefficient restriction
\begin{equation}\begin{array}{lll}
U''(x_0^b) &=& (W_+(x_0^b)\tilde{W}_+(x_0^b))'' \\
&=& W_+''(x_0^b)\tilde{W}_+(x_0^b) +
2W_+'(x_0^b)\tilde{W}_+'(x_0^b) +
W_+(x_0^b)\tilde{W}''_+(x_0^b) \\
&=& 2W_+'(x_0^b)\tilde{W}_+'(x_0^b) = 8\epsilon_0\epsilon_1.
\end{array}\end{equation}

Note, that existence of the fractional powers in the series
expansion leads to undesired poles of $W_0(x)$ at the points
$x_0^b$
\begin{equation}
  W_0(x)^{(\pm)} = \pm {1\over 8} \sqrt{3 U^{(3)}(x_0^b)\over
  \epsilon_0\epsilon_1(x-x_0^b)}+ o(x-x_0^b)^{1/2},
\end{equation}
which can bring the singularity to the potential energy $V_-(x)$.
It is easy to see, that in the case of $U^{(3)}(x_0^b)=0$ the
fractional powers in the series expansions disappear
\begin{equation}\left\{\begin{array}{lll}
  W_+(x) &=& 2\epsilon_0(x-x_0^b)+o(x-x_0^b)^2 \\
  \tilde{W}_+(x) &=& 2\epsilon_1(x-x_0^b)+o(x-x_0^b)^2
\end{array}\right..\end{equation}
The condition $U^{(3)}(x_0^b)=0$ is satisfied in the simplest way
if the point $x_0^b$ is a middle of the periodicity interval and
$U(x)$ is even function with regard to the $x_0^b$, which at the
same time provides the fulfilment of
(\ref{periodicity_condition}).

Then superpotentials read
\begin{equation}\left\{\begin{array}{lll}
  W_0(x)^{(\pm}) &=& B_0^{(\pm)} + o(x-x_0^b) \\
  W_1(x)^{(\pm}) &=& B_1^{(\pm)} + o(x-x_0^b) \\
  W_2(x)^{(\pm}) &=& B_2^{(\pm)} + o(x-x_0^b)
\end{array}\right.,\end{equation}
where
\begin{equation}\left\{\begin{array}{lll}
B_0^{(+)} &=& B_1^{(-)} = B_2^{(+)} = B \\
B_0^{(-)} &=& B_1^{(+)} = B_2^{(-)} = -B \\
B &=& 1/4 \sqrt{32 (\epsilon_0 - \epsilon_1) + U^{(4)}(x_0^b)/(2
\epsilon_0 \epsilon_1)}
\end{array}\right.,\end{equation}
Obtained potential $V_-(x)$ will be continuous function
\begin{equation}\left\{\begin{array}{lll}
  V_-(x)^{(\pm)} &=& \beta_-^{(\pm)} + o(x-x_0^a) \\
  \beta_-^{(\pm)} &=& \epsilon_0 +
  {U^{(4)}(x_0^b) \over 64 \epsilon_0 \epsilon_1} \mp
  {U^{(5)}(x_o^b) \over 320 \epsilon_0 \epsilon_1 B}
\end{array}\right..\end{equation}
Wave functions $\psi_0^-(x)$, $\psi_1^-(x)$, $\psi_2^-(x)$ will
read as follows
\begin{equation}\left\{\begin{array}{lll}
  \psi_0^-(x) &=& 1 + o(x-x_0^b) \\
  \psi_1^-(x) &=& 2\epsilon_0(x-x_0^b) + o(x-x_0^b)^2 \\
  \psi_2^-(x) &=& -2\epsilon_1 + o(x-x_0^b)
\end{array}\right..\end{equation}
Thus, in the vicinity of the second-order zero of the $U(x)$
potential energy $V_-(x)$ and the wave functions $\psi_0^-(x)$,
$\psi_1^-(x)$, $\psi_2^-(x)$ will be continuous functions, if
$x_0^b=x_m$ is the middle of the periodicity interval and $U(x)$
is even function with respect to $x_0^b$. Wave function
$\psi_1^-(x)$ will have node at the points $x_0^b$.

The supersymmetric partner $V_+(x)$ of the $V_-(x)$ potential in
the vicinity of the $x_0^b$ will have the following behavior
\begin{equation}\left\{\begin{array}{lll}
  V_+(x)^{(\pm)} &=& \beta_+^{(\pm)} + o(x-x_0^b) \\
  \beta_+^{(\pm)} &=& \epsilon_0 - 2\epsilon_1 + {U^{(4)}(x_0^b)\over 64\epsilon_0\epsilon_1}
  \pm {U^{(5)}(x_0^b)\over 320 \epsilon_0 \epsilon_1 B
  }
\end{array}\right.,\end{equation}
with the following wave functions
\begin{equation}\left\{\begin{array}{lll}
  \psi_1^+(x) &=& \sqrt{2}\epsilon_0 + o(x-x_0^b) \\
  \psi_2^+(x) &=& 2\sqrt{2}\epsilon_1(\epsilon_0+\epsilon_1)(x-x_0^b) + o(x-x_0^b)^2
\end{array}\right..\end{equation}
Thus, in the vicinity of the second-order zeros $x_0^b$ of the
function $U(x)$ potential energy $V_+(x)$ and the corresponding
wave functions $\psi_1^+(x)$, $\psi_2^+(x)$ will be continuous
function and wave function $\psi_2^+(x)$ will have nodes at the
points $x_0^b$.

Let us analyze the case when the function $U(x)$ has the highest
order of zeros at the points $x_0^c$ using the particular case of
the third-order zeros

\begin{equation}
  U(x)={1\over 6}U^{(3)}(x_o^c)(x-x_o^c)^3+{1\over
  24}U^{(4)}(x_o^c)(x-x_o^c)^4+o(x-x_o^c)^5.
\end{equation}
Then the series expansion for the function $W_+(x)$ in the
vicinity of the points $x_0^c$ will start from the terms which
will be proportional to the $(x-x_0^c)^{3/2}$, thus, condition
(\ref{W_plus_d}) will not be satisfied, and then obtained
potential energy $V_-(x)$ will have poles at the points $x_0^c$.
Consequently, function $U(x)$ should not have zeros of the highest
then second orders.

Singularities at the potential energy, except the zeros of $U(x)$,
can appear at the points where $U'(x)=0$ or $1-\sqrt{R(x)}=0$,
that is
\begin{equation}\left[\begin{array}{lll}
U'(x) &=& 0, \\
U(x) &=& 0, \\
U(x) &=& -2\epsilon_0, \\
U(x) &=& 2\epsilon_1.
\end{array}\right.\end{equation}

Case of $U(x)=0$ was considered in the details above. In the
vicinity of the points $a_0$, where the derivative of $U(x)$ is
equal to zero, i.e. $U'(a_0)=0$ and $U(a_0)\neq 0$, generating
function $U(x)$ can be written as
\begin{equation}
  U(x)=U(a_0)+{1\over 2}U''(a_0)(x-a_0)^2+o(x-a_0)^3.
\end{equation}
Then behavior of function $W_+(x)$ in the vicinity of $a_0$ will
be the following
\begin{equation}
W_+(x)^{(\pm)} = \pm\sqrt{U(a_0)(2\epsilon_0+U(a_0))\over
U(a_0)-2\epsilon_1} + {U''(a_0)\over 4\epsilon_1-2U(a_0)}(x-a_0) +
o(x-a_0)^2,
\end{equation}
in other words, in the vicinity of zeros of $U'(x)$, which do not
coincide with zeros of $U(x)$, obtained solutions will be
continuous functions.

In the vicinity of $b_0$, where $U(b_0)=2\epsilon_1$, function
$W_+(x)$ will behave as follows
\begin{equation}\left\{\begin{array}{lll}
W_+(x)^{(+)} &=& {4\epsilon_1(\epsilon_0+\epsilon_1)\over U'(b_0)}
+o(x-b_0), \\
W_+(x)^{(-)} &=& -{1\over x-b_0}+const+o(x-b_0).
\end{array}\right.\end{equation}
Despite singularity of the function $W_+(x)^{(-)}$, potential
energy and wave functions will be continuous functions, because
pole of $W_+(x)$ satisfies the condition (\ref{W_plus_pole}).

In the vicinity of $c_0$, where $U(c_0)=-2\epsilon_0$, function
$W_+(x)$ can be figured out as follows
\begin{equation}\left\{\begin{array}{lll}
W_+(x)^{(+)} &=& -2\epsilon_0(x-c_0)+o(x-c_0)^2, \\
W_+(x)^{(-)} &=& {U'(c_0)\over 2(\epsilon_0+\epsilon_1)}+o(x-c_0), \\
\end{array}\right.\end{equation}
thus, potential energy and wave functions will be continuous
functions again. Consequently, at the all points, where
denominator of $W_+(x)$ can turn into zero, potential energy
$V_-(x)$ and wave functions $\psi_0^-(x)$, $\psi_1^-(x)$,
$\psi_2^-(x)$ will be free of singularities.

Similar analysis, which we shall omit due to its inconvenience,
with respect to the potential $V_+(x)$ shows, that potential
$V_+(x)$ and corresponding wave functions will be free of
singularities at the all considered points except the points
$c_0$, where potential $V_+(x)^{(+)}$ will have pole with the
following behavior
\begin{equation}
V_+(x)^{(+)} = {1\over (x-c_0)^2}+const+o(x-c_0).
\end{equation}
Fortunately, this singularity can be avoided if within the parts
of periodicity interval which contains $c_0$ we apply solution
$V_+(x)^{(-)}$ instead of $V_+(x)^{(+)}$.

Another way to avoid singularities in the potential $V_+(x)$ is to
exclude zeros in the denominator of $W_+(x)$ by picking up the
amplitude of the function $U(x)$ in such a manner that equations
$U(x)+2\epsilon_0=0$ and $U(x)-2\epsilon_1=0$ not be fulfilled.
Indeed, since energy levels $\epsilon_0$, $\epsilon_1$ are
positively defined values and $U(x)$ is a periodic bounded
function, we can always fit the amplitude of generating function
$U(x)$ using the following rule
\begin{equation}\left\{\begin{array}{lll}
\epsilon_0 < 1/2 \min U(x), \\
\epsilon_1 > 1/2 \max U(x),
\end{array}\right.\end{equation}
where $\min U(x)$ and $\max U(x)$ - minimal and maximal values of
the $U(x)$ at the periodicity interval respectively.

Thus, periodic function $U(x)$ generates quasi exactly solvable
potential $V_-(x)$ with three known eigenfunctions $\psi_0^-(x)$,
$\psi_1^-(x)$, $\psi_2^-(x)$ for the energy values $\epsilon_0 >
0$ and $\epsilon_1 > 0$, if $R(x)\geq 0$ for all periodicity
interval. Simultaneously, function $U(x)$ generates quasi exactly
solvable potential $V_+(x)$ with two known eigenfunctions
$\psi_1^+(x)$, $\psi_2^+(x)$ in the case of
$U(x)\in(-2\epsilon_0;2\epsilon_1)$ and $R(x)\geq 0$ for all
periodicity interval.

To provide free of singularities potential energy and extended
bounded wave functions, $U(x)$ must be even function with respect
to the middle of the periodicity interval $x_m$ and must have
second order zero at this point. Generating function $U(x)$ may
have first-order zeros at the other points of the periodicity
interval and should not have zeros of the highest order. The
derivative of the $U''(x)$ at the point $x_m$ should satisfy the
condition $U''(x_m)=8\epsilon_0\epsilon_1$. It is necessary to use
solutions with opposite signs from the left and right sides with
regard to point $x_m$.

To illustrate the above described method we give a short example.

{\bf Trigonometric extension of the Razavy potential.} Let us
start from the generating function
\begin{equation}
U(x) = 4 \epsilon_0 \epsilon_1 \sin^2 x.
\end{equation}
Similar generating function $U(x) = 4 \epsilon_0 \epsilon_1 \sinh^2 x$ at
$\epsilon_1=\epsilon_0+1/2$ gives well known quasi exactly solvable Razavy
potential \cite{TkKuliy}. Than $\mathfrak{R} (x)$ can be rewritten in the following form
\begin{equation}
\mathfrak{R}(x)=(-1 + 2\epsilon_0 -2\epsilon_1 + 4 \epsilon_0 \epsilon_1 \sin^2 x)
\tan^2 x.
\end{equation}

We shall omit the general expression for the superpotentials and
potential energy as it is huge and rather useless. There are at
least three sets of $\epsilon_0$, $\epsilon_1$, which allow us to
resolve the root in the function $R(x)$ and therefore to
significantly simplify the final results.

The first set is
\begin{equation}
\left\{
\begin{array}{rll}
{4 \epsilon_0 \epsilon_1} & {=} & {0} \cr
{-1 + 2\epsilon_0 - 2\epsilon_1} & {\geq} & {0}
\end{array},
\right.
\end{equation}
for which we obtain trivial solution $\epsilon_0 = 0$ or $\epsilon_1 = 0$,
what leads to the $U(x) = 0$.

In the case of the second set
\begin{equation}
\left\{
\begin{array}{rll}
{-1+2\epsilon_0-2\epsilon_1} & {=} & {-4 \epsilon_0 \epsilon_1} \cr
{-1 + 2\epsilon_0 - 2\epsilon_1} & {\geq} & {0}
\end{array}
\right.
\end{equation}
we obtain $\epsilon_1=-1/2$. Then
\begin{equation}
W_+(x) = {\epsilon_0 \sin 2x \over 1 + \sqrt{2\epsilon_0} \sin x}.
\end{equation}
Function $W_+(x)$ has zeros at the points $x_k=\pi n/2,
n=0,\pm1,...$. The derivations $W_+'(x)$ at this points are
$-2\epsilon_0 / (1 + \sqrt{2\epsilon_0})$ or $2\epsilon_0$ and
condition (\ref{W_plus_d}) is not fulfilled.

The last set
\begin{equation}
\left\{
\begin{array}{rll}
{-1+2\epsilon_0-2\epsilon_1} & {=} & {0} \cr
{4\epsilon_0 \epsilon_1} & {\geq} & {0}
\end{array}
\right.
\end{equation}
gives $\epsilon_1=\epsilon_0-1/2$, then square root can be
rewritten in the following form
\begin{equation}
\left\{
\begin{array}{rll}
{R(x)} & {=} & 2 {\sqrt{\epsilon_0\epsilon_1}\sin x \tan x} \cr
{\epsilon_0} & {\geq} & {1/2}
\end{array}
\right.
\end{equation}

Function $W_+(x)$ reads as follows
\begin{equation}
W_+(x)={2\epsilon_0(\cos^2x+2\epsilon_0\sin^2x)\tan x\over
1+2\sqrt{\epsilon_0\epsilon_1}\sin x\tan x}.
\end{equation}

Function $W_+(x)$ has singularities at the points
$x_k^{(1)}=\pm\arccos \sqrt{\epsilon_0/\epsilon_1} + 2\pi
n,n=0,\pm 1, ...$ and $x_k^{(2)}=\pm\arccos
(-\sqrt{\epsilon_1/\epsilon_0}) + 2 \pi n,n=0,\pm 1, ...$. Due to
the limitation $\epsilon_0\geq 1/2$, solutions $x_k^{(1)}$ belong
to the complex space and thus, can be dismissed. At the points
$x_k^{(2)}$ function $W_+(x)$ has simple poles with the pole
coefficient $-1$, thus potential energy $V_-(x)$ will be regular
function at points $x_k^{(2)}$ for any $\epsilon_0$. Additionally,
function $W_+(x)$ has simple zeros at the points $x_k=\pi n,
n=0,\pm 1, ...$. The derivations $W_+(x)$ at all these points are
equal to $2\epsilon_0$, so all conditions imposed on generating
function $U(x)$ to provide non-singular real potential energy
$V_-(x)$ are satisfied for any $\epsilon_0>1/2$.

Then, using the definition of function $\tilde{W}_+(x)$
(\ref{W_plus_tilde}), solution for superpotentials $W_0(x)$,
$W_1(x)$, $W_2(x)$ (\ref{W_through_Wplus_Wminus}) and relation
between superpotential $W_0(x)$ and potential energy $V_-(x)$
(\ref{V_minus}), we can find three eigenstates of the potential
\begin{equation}
V_-(x)=\epsilon_0-{1\over 2} +
{1\over4}\Big(
\epsilon_0\epsilon_1 - 6\sqrt{\epsilon_0\epsilon_1} \cos x -
\epsilon_0\epsilon_1 \cos 2x \Big),
\end{equation}
where $\epsilon_1 = \epsilon_0 - 1/2$. The energy values of this
eigenstates are $E_0^-=0$, $E_1^-=\epsilon_0$,
$E_2^-=\epsilon_0+\epsilon_1$ and wave functions are given by
(\ref{solution-})

\begin{equation}
\left\{\begin{array}{lll} \psi^-_0(x) &=&
C_0^-e^{\sqrt{4\epsilon_0\epsilon_1}\cos^2{x\over 2}}
\Big(1+4(\sqrt{\epsilon_0\epsilon_1}+\epsilon_1)\cos^2{x\over
2}\Big) \\
\psi_1^-(x) &=& C_1^-e^{\sqrt{4\epsilon_0\epsilon_1}\cos^2{x\over
2}} \epsilon_0 \sin x  \\
\psi_2^-(x) &=& C_2^-e^{\sqrt{4\epsilon_0\epsilon_1}\cos^2{x\over
2}} 2\epsilon_1 \Big(1+
4(\sqrt{\epsilon_0\epsilon_1}-\epsilon_0)\cos^2{x\over 2} \Big)
\end{array}\right..
\end{equation}

\begin{figure}
\centerline{\epsfbox{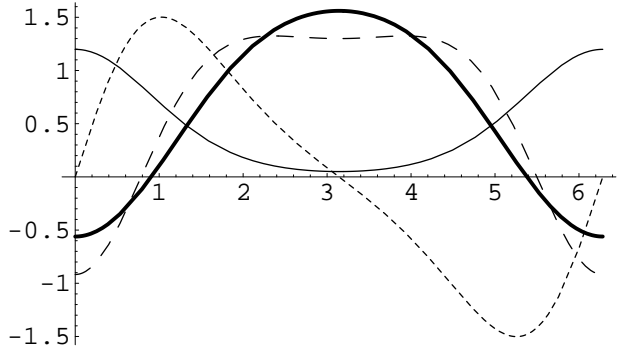}}
\caption{Potential $V_-(x)$ (thick line) and the wave functions $\psi_0^-(x)$,
$\psi_1^-(x)$, $\psi_2^-(x)$ (solid line, short-dashed line and
long-dashed line respectively) at the interval $x\in[0,2\pi]$. Here
$\epsilon_0
= 1$, $C_0^-=0.05$, $C_1^-=0.3$, $C_2^-=1.3$ are used.}
\label{fig1}
\end{figure}

Potential $V_-(x)$ and the wave functions $\psi_0^-(x)$,
$\psi_1^-(x)$, $\psi_2^-(x)$ are presented at the figure
\ref{fig1}.

Because wave function $\psi_0^-(x)$ does not have nodes,
eigenstate with energy $E_0^-=0$ is a ground state of this
potential. The wave functions $\psi_1^-(x)$ and $\psi_2^-(x)$ have
two nodes per interval of periodicity, then eigenstates with
energies $E_1^-$ and $E_2^-$ describe the limits of the second
forbidden energy band.

This quasi exactly solvable potential belongs to the class of QES
potentials presented by Turbiner in his paper \cite{Turbiner} in
the following form
\begin{equation}
V(x)={1\over 2}\Big( -a^2 \cos^2(2\alpha x)-2\alpha a(2n + 1)
\cos(2\alpha x)),
\end{equation}
in the case of $n=1$, $\alpha = 1/2$; $a$ is a free parameter of
quantum mechanics problem.

Now let us consider supersymmetric partner of the potential
$V_-(x)$:
\begin{equation}
V_+(x)={1\over 2}\Big[\epsilon_0^2+{3\over 2}\epsilon_0-1
-\sqrt{\epsilon_0\epsilon_1}\cos x - \epsilon_0\epsilon_1 \cos^2
x\Big] + {\sum\limits_{i=0}^7 a_i \cos^i x\over
2\sum\limits_{i=0}^8 b_i \cos^i x}, \label{Vplus}
\end{equation}

\begin{equation}
\left\{\begin{array}{l}
{a_0 = 16\epsilon_0^2\epsilon_1^2} \\
{a_1=-8\sqrt{\epsilon_0\epsilon_1}\epsilon_0(2-5\epsilon_0+2\epsilon_0^2)} \\
{a_2=-12\epsilon_0(1-2\epsilon_0-2\epsilon_0^2+4\epsilon_0^3)} \\
{a_3=8\sqrt{\epsilon_0\epsilon_1}(1+3\epsilon_0-12\epsilon_0^2+6\epsilon_0^3)} \\
{a_4=1+16\epsilon_0-48\epsilon_0^2(1-\epsilon_0^2)} \\
{a_5=-6\sqrt{\epsilon_0\epsilon_1}(1+2\epsilon_0-12\epsilon_0^2+8\epsilon_0^3)} \\
{a_6=-8\epsilon_1^2\epsilon_0(3+2\epsilon_0)} \\
{a_7=16\epsilon_1^2\epsilon_0\sqrt{\epsilon_0\epsilon_1}}
\end{array}\right.,
\end{equation}

\begin{equation}
\left\{\begin{array}{l}
{b_0=8\epsilon_1\epsilon_0^3}\\
{b_1=8\epsilon_0^2\sqrt{\epsilon_0\epsilon_1}} \\
{b_2=-2\epsilon_0^2(1-12\epsilon_0+16\epsilon_0^2)}\\
{b_3=-8\epsilon_0\sqrt{\epsilon_0\epsilon_1}(3\epsilon_0-1)} \\
{b_4=\epsilon_0(1+10\epsilon_0-48\epsilon_0^2(1-\epsilon_0))} \\
{b_5=2\sqrt{\epsilon_0\epsilon_1}(1-8\epsilon_0+12\epsilon_0^2)} \\
{b_6=-2\epsilon_1^2(-1-4\epsilon_0+16\epsilon_0^2)}\\
{b_7=-8\epsilon_1^2\sqrt{\epsilon_0\epsilon_1}} \\
{b_8=8\epsilon_0\epsilon_1^3.}
\end{array}\right. .
\end{equation}

Since we know eigenfunctions $\psi_1^-(x)$ and $\psi_2^-(x)$ of
Hamiltonian $H_-$, using supersymmetric relations
(\ref{SUSYtransform}) we can find the wave functions $\psi_1^+(x)$
and $\psi_2^+(x)$, which are eigenfunctions of the Hamiltonian
$H_+$ with the corresponding energy values $E_1^+=\epsilon_0$ and
$E_2^+=\epsilon_0+\epsilon_1$:

\begin{equation}
\left\{\begin{array}{l}
{\psi_1^+(x)=C_1^+\epsilon_0
e^{\sqrt{4\epsilon_0\epsilon_1}\cos^2{x\over2}}
{\sum\limits_{i=0}^4 k_i \cos^i x\over
2\sum\limits_{i=0}^4 l_i \cos^i x} }\\
{\psi_2^+(x)=C_2^+e^{\sqrt{4\epsilon_0\epsilon_1}\cos^2{x\over2}}
\sin x {\sum\limits_{i=0}^3 m_i \cos^i x\over 2\sum\limits_{i=0}^4 n_i
\cos^i x}}
\end{array}\right.,
\label{PsiPlus}
\end{equation}
where
\begin{equation}
\begin{array}{ll}

\left\{\begin{array}{l}
{k_0=4\sqrt{2}\epsilon_0\epsilon_1} \\
{k_1=4\sqrt{2\epsilon_0\epsilon_1}} \\
{k_2=-\sqrt{2}(8\epsilon_0\epsilon_1-1)} \\
{k_3=-4\sqrt{2\epsilon_0\epsilon_1}} \\
{k_4=4\sqrt{2}\epsilon_0\epsilon_1} \\
\end{array}\right.

\left\{\begin{array}{l}
{l_0=4\epsilon_0\sqrt{\epsilon_0\epsilon_1}}\\
{l_1=2\epsilon_0} \\
{l_2=2(1-4\epsilon_0)\sqrt{\epsilon_0\epsilon_1}} \\
{l_3=2\epsilon_1} \\
{l_4=4\epsilon_1\sqrt{\epsilon_0\epsilon_1},} \\
\end{array}\right.

\end{array},
\end{equation}

\begin{equation}
\left\{\begin{array}{l}
{m_0=-4\sqrt{2}\epsilon_0\epsilon_1(4\epsilon_0-1)
(\epsilon_1-\sqrt{\epsilon_0\epsilon_1})} \\
{m_1=2\sqrt{2\epsilon_0}(\sqrt{\epsilon_1}-\sqrt{\epsilon_0})
(8\epsilon_0^3-14\epsilon_0^2+7\epsilon_0-1)} \\
{m_2=-\sqrt{2}(\sqrt{\epsilon_0\epsilon_1}-\epsilon_1)
(1-4\epsilon_0-4\epsilon_0^2+16\epsilon_0^3)} \\
{m_3=-4\epsilon_1^2\sqrt{2\epsilon_0}
(\sqrt{\epsilon_1}-\sqrt{\epsilon_0})(4\epsilon_0-1)}
\end{array}\right.,
\end{equation}

\begin{equation}
\left\{\begin{array}{l}
{n_0=2\epsilon_0\sqrt{\epsilon_0\epsilon_1}} \\
{n_1=\epsilon_0} \\
{n_2=-2(\epsilon_0+\epsilon_1)\sqrt{\epsilon_0\epsilon_1}} \\
{n_3=-\epsilon_1} \\
{n_4=2\epsilon_1\sqrt{\epsilon_0\epsilon_1.}}
\end{array}\right.
\end{equation}

\begin{figure}
\centerline{\epsfbox{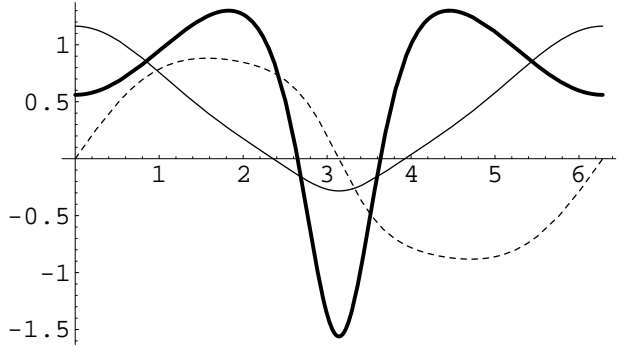}}
\caption{Potential $V_+(x)$ (thick line) and the wave functions $\psi_1^+(x)$,
$\psi_2^+(x)$ (solid line and dashed line respectively) at the
interval $x\in[0,2\pi]$. Here $\epsilon_0
= 1$, $C_1^+=0.2$, $C_2^+=0.7$ are used.}
\label{fig2}
\end{figure}

Thus we obtain QES potential $V_+(x)$ (\ref{Vplus}) with two
exactly know eigenstates $E_1^+ = \epsilon_0$, $\psi_1^+(x)$ and
$E_2^+ = \epsilon_0 + \epsilon_2$, $\psi_2^+(x)$ given by
(\ref{PsiPlus}). Potential $V_+(x)$ and the wave functions
$\psi_1^+(x)$, $\psi_2^+(x)$ are presented at the figure
\ref{fig2}.

Because the wave functions $\psi_1^+(x)$ and $\psi_2^+(x)$ have
two nodes per periodicity interval, the eigenstates with energy
values $E_1^+$ and $E_2^+$ describe limits of the second forbidden
energy band. Note, that QES potential (\ref{Vplus}) does not
belong to the general Turbiner's case \cite{Turbiner} and is
completely new.

\section{Conclusions}
In the present paper we have extended the SUSY method of
constructing well-like QES potentials with three known eigenstates
potentials for the case of periodic potentials.

Thus, periodic function $U(x)$ generates quasi exactly solvable
potential $V_-(x)$ with three known eigenstates $\psi_0^-(x)$,
$\psi_1^-(x)$, $\psi_2^-(x)$ and quasi exactly solvable potential
$V_+(x)$ with two known eigenstates $\psi_1^+(x)$, $\psi_2^+(x)$.

Since we are interested in the real potential energy, condition
$R(x)\geq 0$ should be satisfied. To provide free of singularities
potential energy and extended bounded wave functions, generating
function $U(x)$ must have second order zero at the middle of the
periodicity interval $x_m$ and must be even function with respect
to this point. $U(x)$ may have first-order zeros at the other
points of the periodicity interval and should not have zeros of
the highest order. The derivative of the $U''(x)$ at the point
$x_m$ should satisfy the condition
$U''(x_m)=8\epsilon_0\epsilon_1$. It is necessary to use solutions
for the superpotentials, potentials and wave functions with
opposite signs from the left and right sides with regard to point
$x_m$ to obtain continuous extended wave functions.

As an example of the above described method starting from the
generating functions $U(x)=4\epsilon_0\epsilon_1 \sin^2 x$ we have
obtained QES periodic potential $V_-(x)=\epsilon_0-1/2 +
1/4(\epsilon_0\epsilon_1 - 6\sqrt{\epsilon_0\epsilon_1} \cos x -
\epsilon_0\epsilon_1 \cos 2x)$, which is trigonometric extension
of the well known Razavy QES potential, with three known
eigenstates $E_0^-=0$, $E_1^-=\epsilon_0$ and $E_2^-=\epsilon_0 +
\epsilon_1$, where $\epsilon_1 = \epsilon_0 - 1/2$ and
$\epsilon_0$ is a free parameter. This potential belongs to the
class of QES potentials presented by Turbiner at \cite{Turbiner}.
Eigenstate with energy $E_0^-=0$ is the ground state of this
potential. Eigenstates with energies $E_1^-$ and $E_2^-$ describes
the limits of the second forbidden energy band.

The supersymmetric partner $V_+(x)$ of potential $V_-(x)$ gives us
a new QES periodic potential for which we know two eigenstates
$E_1^+ = \epsilon_0$ and $E_2^+ = \epsilon_0 + \epsilon_2$ in the
explicit form, where $\epsilon_1 = \epsilon_0 - 1/2$ and
$\epsilon_0$ is a free parameter.  This eigenstates describe the
limits of the second forbidden energy band.

Author is grateful to V.~M.~Tkachuk for enlightening suggestions, helpful comments
and discussions.


\begin{thebibliography}{20}
\bibitem{HillsTheorem} Magnus~W and Winkler~S 1966 {\it Hill's equation} (New York:
Winkley)
\bibitem{KronigPenney} de~L~Kronig~R and Penney~W~G 1931 {\it Proc.Roy.Soc.} {\bf 130} 499

\bibitem{Lame} Arscott~F~M 1981 {\it Periodic differential equations} (Oxford: Pergamon)
\bibitem{Ushveridze} Turbiner~A~V and Ushveridze~A~G 1987 {\it Phys.Lett.} A {\bf 126} 181
\bibitem{Turbiner} Turbiner~A~V 1988 {\it Zh.Eksp.Teor.Fiz.} {\bf 94} 33
\bibitem{Turbiner2} Turbiner~A~V 1989 {\it J.Phys.A} {\bf 22} L1
\bibitem{QES1} Brihaye~Y and Godart~M 1993 {\it J.Math.Phys.} {\bf 34} 5283
\bibitem{QHJ} Sree Ranjani~S,Kapoor~A~K and Panigrahi~P~K 2004 {\it
quant-ph/0403196}
\bibitem{Unified} Ramazan Ko\c{c} and Mehmet Koca 2005 {\it quant-ph/0505002} and {\it
quant-ph/0505004}

\bibitem{Witten} Witten~E 1981 {\it Nucl.Phys.Ser.B} {\bf 185} 513
\bibitem{Cooper} Cooper~F, Khare~A and Sukhatme~U 1995 {\it Phys.Rep.} {\bf 251} 267
\bibitem{SUSY1} Jatkar~D~P, Nagaraja Kumar~C and Khare~A 1989 {\it Phys.Lett.A} {\bf 142} 200
\bibitem{SUSY2} Roy~P and Varshni~Y~P 1991 {\it Mod.Phys.Lett.A} {\bf 6} 1257
\bibitem{SUSY3} Gangopadhyaya~A, Khare~A and Sukhatme~U~P 1995 {\it Phys.Lett.A} {\bf 208} 261
\bibitem{QES4} Khare~A and Sukhatme~U 1999 {\it J.Math.Phys.} {\bf 40} 5473
\bibitem{QES5} Sukhatme~U and Khare~A 1999 {\it quant-ph/9902072}
\bibitem{QES6} Khare~A and Sukhatme~U 2001 {\it J.Math.Phys.} {\bf 42} 5652, {\it
quant-ph/0105044}
\bibitem{QES7} Fernandez~C~D~J, Negro~J and Nieto~L~M 2000 {\it Phys.Lett.A} {\bf
275} 338, {\it quant-ph/0301082}
\bibitem{QES9} Khare~A 2001 {\it Phys.Lett.A} {\bf 69} 2888, {\it
quant-ph/0105030}

\bibitem{Tk1} Tkachuk~V~M 1998 {\it Phys.Lett.A} {\bf 245} 177
\bibitem{Tk2} Tkachuk~V~M 1999 {\it  J.Phys.A} {\bf 32} 1291
\bibitem{TkKuliy} Kuliy~T~V and Tkachuk~V~M 1999 {\it  J.Phys.A: Math.Gen.} {\bf 32} 2157
\bibitem{Dolya} Dolya~S~N and Zaslavskii~O~V 2001 {\it J.~Phys.A} {\bf 34} 1981.
\bibitem{Tk3} Tkachuk~V~M 2001 {\it J.Phys.A} {\bf 34} 6339
\bibitem{Tk4} Tkachuk~V~M and Voznyak~O 2002 {\it J.Phys.Stud.} {\bf 6} 40
\bibitem{Tk5} Tkachuk~V~M and Voznyak~O 2002 {\it Phys.Lett.A} {\bf 301} 177

\bibitem{SUSYQM-p1} Dunne~G and Feinberg~G 1998 {\it Phys.Rev.D} {\bf 57} 1271
\bibitem{SUSYQM-p2} Dunne~G and Mannix~J 1998 {\it Phys.Lett.B} {\bf 428} 115

\bibitem{Gendenstein} Gendenshteyn~L~E 1983 {\it Pisma Zh.Eksp.Teor.Fiz.} {\bf 38} 299
\bibitem{deSouza} de Souza Dutra~A 1993 {\it Phys.Rev.A} {\bf 47} R2435
\bibitem{Beckers} Beckers~J, Debergh~N and Nikitin~A~G 1993 {\it Mod.Phys.Lett.A} {\bf
8} 435

\end{thebibliography}
\end{document}